\documentclass[aps,prd,preprint,nofootinbib]{revtex4}
\usepackage{amsfonts}
\usepackage{mathrsfs}
\usepackage{graphicx}
\usepackage{amsmath}
\usepackage{amssymb}
\usepackage{epsfig}
\usepackage{graphicx}
\usepackage{slashed}
\usepackage{color}
\usepackage{multirow}
\usepackage{ulem}
\usepackage{adjustbox}
\usepackage{longtable}
\usepackage{cleveref}

\usepackage{stmaryrd}
\newcommand{\bqa}{\begin{eqnarray}}
	\newcommand{\eqa}{\end{eqnarray}}
\newcommand{\beq}{\begin{equation}}
	\newcommand{\eeq}{\end{equation}}
 
 \newcommand{\IM}{\text{Im}}

\allowdisplaybreaks[1]

\graphicspath{{fig/}{dia/}} \DeclareGraphicsExtensions{.eps}
\begin{document}
	
	\title{Time-reversal asymmetries in $\Lambda_b \to \Lambda(\to p \pi^-)\ell^+\ell^-$  }

	\author { Chao-Qiang Geng,  Chia-Wei Liu, Zheng-Yi Wei\footnote{weizhengyi21@mails.ucas.ac.cn}}
	
	\affiliation{	
		School of Fundamental Physics and Mathematical Sciences, Hangzhou Institute for Advanced Study, UCAS, Hangzhou 310024, China\\
		University of Chinese Academy of Sciences, 100190 Beijing, China}
	\date{\today}

	\begin{abstract}
		We study the decays of $\Lambda_b \to \Lambda(\to p \pi^-) \ell ^+ \ell^-$ with $\ell = (e, \mu, \tau)$.
In particular, we examine the full angular distributions with polarized $\Lambda_b$ and identify the time-reversal asymmetries
or T-odd observables. 
By using the homogeneous bag model, we find that the decay branching fractions of $\Lambda_b \to \Lambda \ell^+\ell^-$ are 
$(9.1\pm 2.5,7.9\pm 1.8, 2.1\pm 0.2)\times 10^{-7}$ for $\ell =(e, \mu ,\tau)$, respectively.  In addition,
we obtain that  $A_{FB}^{\ell} = -0.369\pm 0.007$ and $A_{FB}^{h}=-0.333\pm 0.004$, averaged in the range of $15 \leq q^2 \leq 20~ \text{GeV}^2$. 
These results are well consistent with the current experimental data. 
We also explore  the T-odd observables in  $\Lambda_b \to \Lambda(\to p \pi^-) \mu^+ \mu^-$, which are sensitive to new physics~(NP).
%With the complex Wilson coefficients, the experimental central value of $K_{10}$ at LHCb can be explained. 
Explicitly, we illustrate that the current experimental measurement  from one of the T-odd observables favors the existence of  NP, such as the extra $Z$-boson model.
%,  which couples  to the left-handed  quarks and leptons. 
% From the experimental data of $\Lambda_b \to \Lambda(\to p\pi^-) \mu^+ \mu^-$, 
	\end{abstract}
	
	\maketitle

	\section{Introductions}

The CP violating observables  in $b\to s \ell^+ \ell^-$ with $\ell= (e,\mu, \tau)$ play  important roles to search for  new physics~(NP) as they are highly suppressed in the standard model~(SM)~\cite{Bobeth:2011gi,Kruger:1999xa,Altmannshofer:2008dz,Bobeth:2008ij,LHCb:2017slr,Fleischer:2017ltw,Kindra:2018ayz}. 
 In recent years, special attentions have been given to the decays of $B \to K^{(*)} \mu^+ \mu^- $ and $B_{s}\to \phi \mu^+ \mu^- $~\cite{LHCb:2013tgx,LHCb:2014cxe,LHCb:2013ghj}. Benefited by the experimental developments, precise measurements of  the angular observables are now accessible~\cite{CMS:2015bcy,ATLAS:2018gqc,LHCb:2020lmf,LHCb:2021xxq,LHCb:2020gog,CMS:2017rzx,LHCb:exp,LHCb:2015svh,LHCb:2018angular}. 
 These observables are useful in disentangling the  helicities, providing reliable methods to probe the  Lorentz structure of NP ~\cite{Buchalla:1995vs,Mott:2011cx,Roy:2017dum,Das:2018iap,Aliev:2002nv,Huang:1998ek,gutsche,Boer:2014kda}.
Besides, the ratios of $R_{K^{(*)}} \equiv \Gamma(B \to K^{(*)} \mu^+ \mu^-)/\Gamma(B \to K^{(*)} e^+ e^-)$ were measured, where  discrepancies against the SM were given. In particular, 3.1$\sigma$ and 2.5$\sigma$ deviations have been found in
  $R_K(1.1 \text{GeV}^2\leq q^2 \leq6.0\text{GeV}^2)$ and $R_{K^*}(0.045 \text{GeV}^2\leq q^2 \leq6.0\text{GeV}^2)$ ~\cite{LHCb:2021trn,LHCb:2017avl},
 showing that the lepton universality may be violated by NP.
Very recently, a global fit of $b\to s \ell^+ \ell^-$ with the $B$ meson experiments has been performed~\cite{SinghChundawat:2022zdf}, and the large complex Wilson coefficients have been demonstrated to be permitted by the current experimental data.

The baryonic decays of $\Lambda_b \to \Lambda(\to p \pi^-) \ell^+ \ell^-$ are  interesting for several reasons. For polarized $\Lambda_b$, 
the decays
of $\Lambda_b \to \Lambda (\to p \pi^-) \ell^+ \ell^-$
 provide  dozens of angular observables, which are three times more than those in  $B \to K \mu^+ \mu^-$. 
The  polarization fraction~$(P_b)$ of $\Lambda_b$ is reported as $(6\pm 7)\%$  at the center of mass energy 7 TeV of $pp$ collisions~\cite{LHCb:2013hzx}.
The full angular distribution of  $\Lambda_b \to \Lambda(\to p \pi^-) \mu^+ \mu^-$  has been  measured  at LHCb~\cite{LHCb:2018angular}. 
Notably, the 
experiment obtains that one of the physical observables is given by
\begin{equation}
K_{10} = -0.045\pm 0.037 \pm 0.006 \,,
\end{equation}
which  deviates to the SM prediction of $K_{10} \approx 0$ by  $1.2\sigma$. 
It is reasonable to expect that the precision will be improved in the forthcoming update. 
In this work, we will show explicitly that $K_{10}$ is an T-odd quantity, which can be sizable in the presence of NP.  

On the theoretical aspect, 
the angular distributions of $ \Lambda_b \to \Lambda \mu^+ \mu^-$ have been studied intensively~\cite{gutsche,Boer:2014kda,Blake:2017angular}. In particular,
an analysis of NP with real Wilson coefficients has been performed in Ref.~\cite{Blake:2019guk}, in which  $P_b=(0\pm 5)\%$ is found at $1\sigma$ confidence level. 
In this work,
we would like to focus on the time-reversal~(T) violating observables induced by the complex NP Wilson coefficients.
In comparison to the CP violating quantities, the T violating ones do not require strong phases.  In the leptonic decays, this feature is very useful as strong phases are  often negligible. 

This paper is organized as follows. In Sec.~II , we decompose $\Lambda_b\to \Lambda \ell^+\ell^-$ into products of two-body decays.  In Sec.~III , we  construct T-odd observables. In Sec.~VI, we briefly review the  angular distributions of $\Lambda_b\to\Lambda(\to p \pi^-) \ell^+ \ell^-$,  and identify  the T-odd observables. In Sec.~V, we give the numerical results from the homogeneous bag model~(HBM).  We conclude the study in Sec.~VI. 
	
	\section{Helicity amplitudes}
The amplitudes of 
	 $ \Lambda_b \to  \Lambda \ell^+ \ell^- $, induced by the  transitions of $b\to s\ell^+ \ell^-$   at the quark level,
 are given as~{\cite{Buchalla:2000sk}} 
	\begin{equation}
	    \label{eq1}
     \begin{aligned}
		\frac{G_F}{\sqrt 2} \frac{\alpha V^{*}_{ts} V_{tb}}{2\pi }  \left[ \langle \Lambda|\bar{s}
j_1^\mu 
b 
|\Lambda_b\rangle \bar{\ell}\gamma_{\mu} \ell + \langle \Lambda|\bar{s} j_2^\mu  b  |\Lambda_b\rangle \bar{\ell} \gamma_ {\mu} \gamma_5 \ell \right] ,
	\end{aligned}
    \end{equation}
	where $G_F$  is the Fermi constant,  
	$V_{ts, tb}$ are the Cabibbo-Kobayashi-Maskawa~(CKM) matrix elements, 
\begin{equation}
\begin{aligned} \label{eq3}
&j_1^\mu = ( C_9^{eff}+C^{\text{NP}}_9) L^\mu   -\frac{2 m_{b}}{q^2}C_{7\gamma}^{eff}
 i\sigma^{\mu q}(1+\gamma_5)   + (C_{L}+C_{R})R^{\mu} \,, \\
&j_2^\mu = (C_{10} + C^{\text{NP}}_{10}) L^ \mu  + (C_{R}-C_{L})R^{\mu} \,,
\end{aligned}
\end{equation}
$C^{(eff)}$ are the (effective) Wilson coefficients,
$\sigma^{\mu q} =i(\gamma^{\mu}\gamma^{\nu}-\gamma^{\nu}\gamma^{\mu})q_{\nu}/2$  with $q = (q^0 , \vec{q}~)$ the four-momentum of $\ell^+ \ell^-$,  $L^\mu = \gamma^{\mu}(1-\gamma_5)$,  $R^\mu = \gamma^{\mu}(1+\gamma_5)$, and  $m_{q}$ stands for the quark mass. 
The first (second) term in Eq.~\eqref{eq1} can be interpreted as $\Lambda_b \to \Lambda j_{eff}^{1(2)}$ followed by $j_{eff}^{1(2)}\to \ell^+ \ell^-$, where $j_{eff}^{1(2)}$ is an  effective off-shell (axial) vector boson,
conserving the parity in its cascade decays, and $j^\mu_{1,2}$ are the couplings of $b-s-j_{eff}^{1,2}$. 
Alternatively,
the interpretation can also be rephrased as $\Lambda_b\to \Lambda j_{eff}^{R,L}$, where $j_{eff}^{R(L)}$ couples only to the right-handed (left-handed) leptons, given as 
	\begin{equation}
	    \label{eq4}
     \begin{aligned}
		\frac{G_F}{\sqrt 2} \frac{\alpha V^{*}_{ts} V_{tb}}{2 \pi }  \left[ \langle \Lambda|\bar{s}
j_+^\mu 
b 
|\Lambda_b\rangle \bar{\ell}R_\mu\ell + \langle \Lambda|\bar{s} j_-^\mu  b  |\Lambda_b\rangle \bar{\ell} L_\mu  \ell \right] ,
	\end{aligned}
    \end{equation}
where $j_\pm^\mu= (j_1^\mu \pm j^\mu _2)/2 $. 
In the SM,  $C^{\text{NP}}_{9,10} = C_{L,R}=0$ and the others are~\cite{gutsche,faustov}
  \begin{equation}
        \begin{aligned}
C_{7\gamma}^{eff}&=-0.313,         
        \\
        C_{9}^{eff} &= C_{9} + h(\frac{m_c}{m_b} ,\frac{q^2}{m_b ^2})  -\frac{1}{2} h(1,\frac{q^2}{m_b ^2})(4C_{3} +4C_4 +3C_5 +C_6) \\&-\frac{1}{2} h(0,\frac{q^2}{m_b ^2})(C_{3} +3C_4 ) +\frac{2}{9} (3C_3 +C_4 +3C_5 +C_6),
        \end{aligned}
    \end{equation}
where
    \begin{equation}
        \begin{aligned}
        \begin{split}
        h(\frac{m_c}{m_b} ,\frac{q^2}{m_b}) &= -\frac{8}{9}\ln \frac{m_c}{m_b}
                +\frac{8}{27} +\frac{4}{9}x -\frac{2}{9}(2+x) \\
                &\times |1-x|^{1/2} \left\{    \begin{aligned}    \left(\ln \big| \frac{\sqrt{1-x}+1}{\sqrt{1-x}-1}\big| -i \pi \right), &~~ \text{for}~~  x\ < 1 ,\\ 
                2 \text{arctan}\frac{1}{\sqrt{x-1}} , & ~~\text{for}~~  x > 1 ,  \end{aligned}  \right.         \\
        h(0,\frac{q^2}{m_b}) &= \frac{8}{27}  -\frac{4}{9}\ln \frac{q^2}{m_b} + \frac{4}{9}i\pi ,
        \end{split}
        \end{aligned}
    \end{equation}
and $x = 4m_c^2/q^2$.
Their explicit values can be found  in Ref.~\cite{faustov}.	

As the parity is conserved in $j_{eff}^{1,2} \to \ell^+ \ell^-$, it is easier to obtain the angular distributions with the $j_{eff}^{1,2}$ interpretations. However, to examine NP, the second interpretation with  $j_{eff}^{R,L}$ is more preferable as NP is likely to couple with the leptons with the same handedness.
We note that physical quantities are of course independent of the interpretations. For our purpose, the angular distributions are studied with   $j_{eff}^{1,2}$, whereas NP  with  $j_{eff}^{R,L}$.

By decomposing the Minkowski metric as \begin{equation}
g^{\mu \nu } = \epsilon_t^\mu \epsilon_t^{*\nu} - \sum_{\lambda=0,\pm} \epsilon_\lambda^\mu \epsilon_\lambda^{*\nu}\,,
\end{equation}
 we arrive at
\begin{equation}\label{good}
		\frac{G_F}{\sqrt 2} \frac{\alpha V^{\dagger}_{ts} V_{tb}}{2\pi }
		\sum_{m=1,2}  \left(
		L_t^m B_t^m    - \sum_{\lambda=0,\pm}  L_{\lambda}^m B_{\lambda} ^m 
		\right)\,,
\end{equation}
where
\begin{equation}\label{good2}
            \begin{aligned}
&B_{\lambda_m} ^m  =  \epsilon_{\lambda_m}  ^{\ast\mu} \langle \Lambda  | \bar{s} j_m^\mu b | \Lambda_b \rangle\,,&~~~L_{\lambda_m}^1  = \epsilon_{\lambda_m}^{\mu} \bar{u}_{\ell}  \gamma_ {\mu} v  \,,&~~~L_{\lambda_m}^2  = \epsilon_{\lambda_m}^{\mu} \bar{u}_{\ell} {\gamma_ {\mu}\gamma_5} v  \,,
         \end{aligned}
	\end{equation}
 ${\lambda_m}  = (t, 0 ,\pm)$ is the helicity of $j_{eff}^m$ with  $t$ indicating spin-0 off-shell contributions, and $\epsilon$ are the polarization vectors of $j_{eff}^m$, given as~\cite{TRSEMI} 
 \begin{equation}\label{restPolar}
		{\epsilon} ^\mu _\pm  = \frac{1}{\sqrt{2}}(0,  \pm 1, i , 0 )^T\,, \quad {\epsilon}_0^\mu  = (0 ,0,0,-1 )^T \,, \quad  {\epsilon}^\mu  _t = (  -1 ,0,0,0 )^T\,,
\end{equation}
and 
\begin{equation}\label{qframe} 
		{\epsilon} ^\mu _\pm  = \frac{1}{\sqrt{2}}(0,  \mp 1, i , 0 )^T\,, \quad {\epsilon}_0^\mu  = \frac{1}{\sqrt{q^2}}(|\vec{q}\,| ,0,0,-q^0 )^T \,, \quad  {\epsilon}^\mu  _t =-\frac{1}{\sqrt{q^2}} q^\mu ,
	\end{equation}
in the center of mass~(CM) frames of $j_{eff}^m$ and $\Lambda_b$, respectively.
In  Eq.~\eqref{good},
the amplitudes are decomposed as the products of  Lorentz scalars, 
where $B_{\lambda_m}$ and $L_{\lambda_m}$
describe $\Lambda _b \to \Lambda  j_{eff}^m$ and $j_{eff}^m \to \ell^+ \ell^-$, respectively,  reducing the three-body problems to two-body ones.

To deal with the spins, we adopt the helicity approach. The projection operators in the $SO(3)$ rotational~$(SO(3)_R)$ group are given by
\begin{equation}\label{pro}
 |J, M\rangle \langle J,  N| =\frac{2J+1}{8 \pi^2 }\int d\phi d\theta d \psi R_z(\phi) R_y(\theta) R_z(\psi) D^{J\dagger} (\phi,\theta,\psi)^N\,_M \,,
\end{equation}
where $N$ and $M$ are the angular momenta toward the $\hat{z}$ direction, the Wigner-$D$ matrices are defined by
\begin{equation}\label{Dmatrix}
D^J(\phi,\theta,\psi) ^{M}\,_N \left\langle
J,N|J,N\right \rangle = 
\left \langle J, M\left | R_z(\phi) R_y(\theta) R_z(\psi) \right| J,N \right \rangle\,,
\end{equation}
and $R_{y(z)}$ are the rotation operator pointing toward $\hat{y}(\hat{z})$. 
We note that it is important for Eq.~\eqref{pro} to be a linear superposition of  $R_{y,z}$, which commutes with  scalar operators. 
In the following, we take the shorthand notation of $D^J(\phi,\theta)\equiv D^J(\phi,\theta,0)$. 

The simplest two-particle state with a nonzero momentum is defined by 
\begin{equation}\label{2pstate}
|p\hat{z}, \lambda_1,\lambda_2 \rangle \equiv 
L_z|\vec{p}= 0 , J_z = \lambda_1 \rangle_1 \otimes  L_z' |\vec{p}=0, J_z= - \lambda_2 \rangle_2\,,
\end{equation}
where $\lambda_{1,2}$ are the helicities, the subscripts denote the particles, and   $L^{(\prime)}_z$ is the Lorentz boost, which brings the first (second) particle to $(-)p\hat{z}$.
As $L_z^{(\prime)}$ commutes with $R_z$, the state defined by Eq.~\eqref{2pstate} is an eigenstate of $J_z = \lambda_1-\lambda_2$. 
Plugging Eq.~\eqref{pro} into Eq.~\eqref{2pstate}  with  $N= \lambda_1-\lambda_2$, we arrive at 
\begin{equation}\label{ptoJ}
\begin{aligned}
&|\vec{p}\,^2, \lambda_1,\lambda_2 ;J,J_z\rangle = \frac{2J+1 }{4\pi} \int d\phi d\cos\theta R_z(\phi) R_y(\theta) |p\hat{z}, \lambda_1,\lambda_2\rangle_{1,2} D^{J*}(\phi,\theta)^{J_z}\,_{N}\,,
\end{aligned}
\end{equation}
which expresses the angular momentum eigenstate as the linear superposition of the three-momentum ones. 
Conversely, we have 
\begin{equation}\label{Jtop}
| p \hat{z}, \lambda_1 ,\lambda_2 \rangle = \sum_{J} | \vec{p}\,^2, 
\lambda_1 ,\lambda_2;
J,N \rangle\,. 
\end{equation}
Note that the identities of Eqs.~\eqref{ptoJ} and \eqref{Jtop} purely come from the mathematical consideration.
The simplification happens when the angular momentum conservation is considered.
At the CM frames of  $\Lambda_b$ and $j_{eff}^{m}$, 
it is clear that only $J=1/2$ and $J=(0,1)$ need to be considered for the $ \Lambda j_{eff}^{m}$  and $\ell^+\ell^-$ systems, respectively.
% since the $\Lambda_b$ and $j_{eff}^{1(2)}$ are spin-half and  spin-one, respectively. 

Utilizing Eq.~\eqref{Jtop}, we have that 
\begin{equation}\label{17}
\langle \vec{p}\,^2 ,\lambda_1,\lambda_2 ; J, N | {\cal S} |  J , J_z;i \rangle  = \langle p\hat{z},\lambda_1,\lambda_2 |{\cal S}|  J , J_z;i \rangle\,,
\end{equation}
where 
${\cal S}$ is an arbitrary scalar operator, and $|J , J_z;i\rangle$ stands for an arbitrary initial state. 
In Eq.~\eqref{17},  the final state in the left side possesses a definite angular momentum, which is irreducible under  $SO(3)_R$, {\it i.e.} it contains only the dynamical details. 
On the contrary, the one in the right side is  three-momentum eigenstate,  containing less physical insights but providing a way to compute the helicity amplitude.

Let us return to  $\Lambda_b \to \Lambda j_{eff}^{m}$ and 
$j_{eff}^{m} \to \ell^+ \ell^-$. 
We take the uppercase and lowercase of $H$ and $h$
for the helicity amplitudes of $\Lambda_b \to \Lambda j_{eff}^{m}$ and 
$j_{eff}^{m} \to \ell^+ \ell^-$, respectively. To be explicit,  we have  
\begin{equation}
        \label{h+-}
	    \begin{aligned}
		H_{\lambda_\Lambda \lambda_m}^m
		&= B_{\lambda_m}\left(
		\lambda_{\Lambda_b}  = \lambda_{\Lambda} - \lambda_m, \lambda_{\Lambda} , \vec{p}_{\Lambda} = -\vec{q} = |\vec{p}_{\Lambda} | \hat{z}
		\right)\,,\\
h_{0,\lambda_+ \lambda_-}^{m }  &= L_{t}^m(\lambda_+, \lambda_-, \vec{q}=0, \vec{p}_+ = -\vec{p}_- =| \vec{p}_+| \hat{z} )\,, \\
h_{1, \lambda_+ \lambda_-}^{m }  &= L_{\lambda_+ - \lambda_- }^m( \lambda_+,  \lambda_-, \vec{q}=0, \vec{p}_+ = -\vec{p}_- =| \vec{p}_+| \hat{z} )\,, 
\end{aligned}
\end{equation}
where  $\lambda_{\Lambda_b} $ corresponds to the angular momentum of $\Lambda_{b }$,  
$(\lambda_{\Lambda},\lambda_\pm )$  are the helicities of $(\Lambda, \ell^\pm) $, and 
	$\vec{p}_{\Lambda}$ and  $\vec{p}_\pm  $ are the 3-momentua of $\Lambda$ and $\ell^\pm $ in the CM frame of $\Lambda_ b$ and $j_{eff}^m$, respectively.
Theoretically speaking, the dynamical parts of the amplitudes are extracted by Eq.~\eqref{17}, whereas the kinematic dependencies are governed by $D^J$.

 For compactness, we take the abbreviations 
\begin{equation}
\begin{aligned}
&|a^{m}_\pm \rangle = |\vec{p}\,^2, \pm 1/2,0; J,J_z\rangle,  &|b^{m}_\pm\rangle = |\vec{p}\,^2, \mp 1/2,\mp 1 ; J,J_z\rangle,~~~& |c^{m}_\pm \rangle = |\vec{p}\,^2, \pm 1/2,t; J,J_z\rangle~~~ \\
&a^{m}_{\pm} = H^{m}_{\pm \frac{1}{2}0} = \langle a_\pm ^m | {\cal S}_{eff}|\Lambda_b\rangle , 
 & b^{m}_{\pm}= H^{m}_{\mp \frac{1}{2}\mp1} =  \langle a_\pm | {\cal S}_{eff} |\Lambda_b\rangle ,&~~c^{m}_{\pm} = H^{m}_{\pm \frac{1}{2}t} = \langle c_\pm ^m | {\cal S}_{eff}|\Lambda_b\rangle ,
\end{aligned}
\end{equation}
where ${\cal S}_{eff}$ is the  transition operator responsible for $\Lambda_b \to \Lambda j_{eff}^m$, and $J_z$ is not written down explicitly.  The artificial ${\cal S}_{eff}$ is needed to interpret  $\Lambda_b\to \Lambda \ell^+ \ell^- $ as  products of two-body ones. For the $\Lambda_b\to \Lambda j_{eff}^{R,L}$ interpretation, the helicity amplitudes are 
\begin{eqnarray}
&&a_\pm^R  = \frac{1}{\sqrt{2}}(a_\pm^1 + a_\pm ^2)\,,~~~ a_\pm^L =  \frac{1}{\sqrt{2}} (a_\pm^1 - a_\pm^ 2)\,,\nonumber\\
&&b_\pm^R  = \frac{1}{\sqrt{2}}(b_\pm^1 + b_\pm ^2)\,,~~~~ b_\pm^L =  \frac{1}{\sqrt{2}} (b_\pm^1 - b_\pm ^2)\,,\nonumber\\
&&c_\pm^R  = \frac{1}{\sqrt{2}}(c_\pm^1 + c_\pm ^2)\,,~~~~ c_\pm^L =  \frac{1}{\sqrt{2}} (c_\pm^1 - c_\pm ^2)\,. 
\end{eqnarray}

\section{T-odd observables}

From Eq.~\eqref{eq3}, we see that the NP contributions are absorbed into the couplings of $b-s-j_{eff}^r$, while the Lorentz structures of $j_{eff}^r \to \ell^+ \ell^-$ are simple with $r=(1,2,R,L)$. Thus, to discuss the NP effects, it is sufficient to study $\Lambda_b \to \Lambda j_{eff}^r$.

 The most simple T-odd operator in $\Lambda_b \to \Lambda j_{eff}^m$ is defined as ~\cite{TRlamV}
 \begin{equation}\label{Todd}
 \hat{T} = (\vec{s}_{\Lambda} \times \vec{s}_{m} )\cdot \hat{p}_{\Lambda},   
 \end{equation}
 $\vec{s}_{\Lambda}$ and $\vec{s}_{m}$ are the spin operators of $\Lambda$ and $j^{m}_{eff} $, respectively, and  $\hat{p}_{\Lambda}$ is the  unit vector of $\vec{p}_{\Lambda}$. The spin operators can only be defined for the massive objects, given as
 \begin{equation}
M\vec{s} = P^0 \vec{J} - \vec{p} \times \vec{K } - \frac{1}{P^0 + M }\vec{p}(\vec{p}\cdot \vec{J})\,,
 \end{equation}
 where $M$ is the particle mass, and $P^0$, $\vec{p}$, $\vec{J}$ and $\vec{K}$ are the time translation,
space translation, rotation 
 and Lorentz boost generators, respectively. 
As $(\vec{p}, \vec{J})$ and $\vec{K}$ are T-odd and T-even, respectively,  $\vec{s}$ is T-odd. In addition, $\vec{s}$ satisfies the relations 
\begin{equation}\label{eq21}
\begin{aligned}
& \vec{s} \cdot \vec{p} = \vec{J} \cdot \vec{p} \,,~~~[s_i,s_j] = i \epsilon^{ijk} \epsilon_k\,,~~~[s_i,p_j] =0 \,,  \\
&\vec{s} \exp(i\vec{K}\cdot \vec{\omega})|\vec{p}=0, J_z = M\rangle =  \exp(i\vec{K}\cdot \vec{\omega}) \vec{J} |\vec{p}=0,J_z =M\rangle\,, 
\end{aligned}
\end{equation}
with arbitrary $\vec{\omega}$. The key of solving the eigenstates of $\hat{T}$ relies on  that $\hat{T}$ is a scalar operator. We have 
\begin{equation}\label{ptoJT}
\begin{aligned}
&\hat{T}|\vec{p}\,^2, \lambda_1,\lambda_2 ;J,J_z\rangle \\&~~~= \frac{2J+1 }{4\pi} \int d\phi d\cos\theta R_z(\phi) R_y(\theta) \hat{T} |p\hat{z}, \lambda_1,\lambda_2\rangle_{1,2} D^{J*}(\phi,\theta)^{J_z}\,_{\lambda_1-\lambda_2}\,,
\end{aligned}
\end{equation}
and 
\begin{equation}
\hat{T} |p \hat{z}, \lambda_1, \lambda_2 \rangle 
= \frac{i}{2}
(s_\Lambda^+ s_m^- - s_\Lambda^- s_m^+) 
|p \hat{z}, \lambda_1, \lambda_2 \rangle \,,
\end{equation}
with $s^\pm = s_x \pm  is_y$. 
It is then straightforward to show that
\begin{equation}
    \hat{T} |a^{m}_\pm \rangle =\pm \frac{i}{\sqrt{2}}|b^{m}_\pm\rangle ,~~~
    \hat{T} |b^{m}_\pm \rangle = \mp \frac{i}{\sqrt{2}}|a^{m}_\pm \rangle ,
\end{equation}
resulting in the eigenstates
\begin{equation}
\begin{aligned}
&|\lambda_{T}^m=\pm\frac{1}{\sqrt{2}},\lambda_{\text{tot}} = \frac{1}{2} \rangle = \frac{1}{\sqrt{2}}(|a^{m}_+ \rangle \mp i|b^{m}_+ \rangle),\\
&|\lambda_{T}^m=\pm\frac{1}{\sqrt{2}},\lambda_{\text{tot}} = -\frac{1}{2} \rangle = \frac{1}{\sqrt{2}}(|a^{m}_- \rangle \pm i|b^{m}_- \rangle)\,,
\end{aligned}
\end{equation}
where 
$\lambda_T^m$ and 
$\lambda_{\text{tot}}$ are the eigenvalues of $\hat{T}$ and $\vec{J}\cdot \vec{p}$, respectively. 
They are also the eigenstates of $\vec{J}\cdot \vec{p}$, as $\hat{T}$ commutes with both $\vec{J}$ and $\vec{p}$.
Note that $c_{\pm}^{m}$ are not involved since they are contributed by  spinless $j_{eff}^m$.

Because $ \hat{T} $ and $\vec{J}\cdot\vec{p}$ are  T-odd  and T-even, respectively, we have
\begin{equation}\label{27}
{\cal I}_t |\lambda_T^m, \lambda_{\text{tot}}\rangle  =e^{i\theta_T} |-\lambda_T^m,  \lambda_{\text{tot}}\rangle\,,~~~{\cal I}_s |\lambda_T^m, \lambda_{\text{tot}}\rangle  =e^{i\theta_m} |-\lambda_T^m,  -\lambda_{\text{tot}}\rangle\,,
\end{equation}
where ${\cal I}_{t(s)}$ is the time-reversal (space-inversion) operator, and $\theta_{T,m}$ depend on the conventions. 
On the other hand, ${\cal I}_s$ would interchange $j_{eff}^R$ and $j_{eff}^L$, given as 
\begin{equation}
{\cal I}_s |\lambda_T^R, \lambda_{\text{tot}}\rangle  =e^{i\theta_R} |-\lambda_T^L,  -\lambda_{\text{tot}}\rangle\,,~~~
{\cal I}_s |\lambda_T^L, \lambda_{\text{tot}}\rangle  =e^{-i\theta_R} |-\lambda_T^R,  -\lambda_{\text{tot}}\rangle\,,
\end{equation}
with
\begin{equation} \small 
|\lambda_T^R , \lambda_{\text{tot}} \rangle = \frac{1}{\sqrt{2}}\left(
|\lambda_T^1 , \lambda_{\text{tot}} \rangle  +  |\lambda_T^2 , \lambda_{\text{tot}} \rangle
\right)\,,~~~|\lambda_T^L , \lambda_{\text{tot}} \rangle = \frac{1}{\sqrt{2}}\left(
|\lambda_T^1 , \lambda_{\text{tot}} \rangle  -  |\lambda_T^2 , \lambda_{\text{tot}} \rangle
\right)\,,
\end{equation}
since $j_{eff}^1$ and $j_{eff}^2$ have opposite parity.

For each combinations of $\lambda_{\text{tot}} $ and $j_{eff}^r$, we define an T-odd quantity 
\begin{equation}\label{28}
{\cal T}_{\lambda_{\text{tot}}}^{\,r} \equiv 
|\langle \lambda_T^r =1/\sqrt{2},\lambda_{\text{tot}}  | {\cal S}_{eff} | \lambda_b \rangle|^2 - |\langle \lambda_T^r =-1/\sqrt{2},\lambda_{\text{tot}}  | {\cal S}_{eff} | \lambda_b \rangle|^2\,,
\end{equation}
which vanishes if ${\cal S}_{eff}$ is invariant under ${\cal I}_t$.
Explicitly, we find 
\begin{equation}
\label{naive t-odd}
\begin{aligned}
&
{\cal T}_{+} ^{\,r} 
=- 2 \text{Im} \left (a_+^r \overline{b_+^r} \right ) \,,~~
&{\cal T}_{-} ^{\,r} = 2 \text{Im} \left (a_-^r \overline{b_-^r} \right )\,,
\end{aligned}
\end{equation}
% checked
which are proportional to the relative complex phase. 
They are called as T-odd quantities as ${\cal I}_t$ interchanges the final states of  the two terms in Eq.~\eqref{28}. 

The operator of  $\hat{T}$ contains  $\vec{s}_\Lambda$, which is difficult to be measured directly. To probe the spin of $\Lambda$, it is plausible to study the cascade decays of $\Lambda \to p \pi^- $. Subsequently, the final states involve four particles $p \pi^- \ell^+ \ell^-$, containing three independent three-momenta. It is then possible to observe the triple product
\begin{equation}\label{30}
\alpha  ( \vec{p}_+\times \vec{p}_p ) \cdot \vec{p}_\Lambda,    
\end{equation}
 where $\alpha$ is the polarization asymmetry in $\Lambda \to p \pi^-$, and $\vec{p}_p$ is the three-momentum of the proton.  Notice that $\alpha$ is a necessary component in Eq.~\eqref{30} as $\vec{s}_\Lambda$ does not affect $\vec{p}_p$ if  $\alpha = 0$. Observe that Eq.~\eqref{30} is P-even. Therefore, we have to construct P-even observables out of Eq.~\eqref{naive t-odd}. From the transformation rules, it is easy to see that
%  \begin{equation}\label{34}
% {\cal T} ^{R} \equiv {\cal T}_\pm ^{R} - {\cal T}_\mp^{L}\,,~~~{\cal T} ^{L} \equiv {\cal T}_\pm ^{L} - {\cal T}_\mp^{R}\,,    
%  \end{equation}
 \begin{equation}\label{34}
{\cal T} ^{R} \equiv {\cal T}_- ^{R} - {\cal T}_+^{L}\,,~~~{\cal T} ^{L} \equiv  {\cal T}_-^{L} - {\cal T}_+ ^{R}  \,,    
 \end{equation}
which are both T-odd and P-even.

\section{Angular distributions}
The lepton helicity amplitudes are calculated  as 
	\begin{equation}\label{ratiosof Lepton}
        \begin{aligned}
		h_{0,++}^1 &= 0 \,,~~~ &&h_{1,++}^1 = 2M_\ell \,, \\
        h_{0,++}^2 &= 2M_\ell \,,~~~&&h_{1,++}^2 = 0 \,,\\
        h_{1,+-}^1 &= -\sqrt{2q^2} \,, ~~~&&h_{1,+-}^2 = \sqrt{2q^2(1-\delta_\ell)}\,,
        \end{aligned}
	\end{equation}
where $\delta_\ell = 4M_\ell^2 / q^2$ and $M_\ell$ is the lepton mass.
On the other hand, 
the  baryonic matrix elements are conventionally parameterized by the form factors, given by
\begin{equation}
	    \begin{aligned}
	 \label{eqformfactor}
  \langle \Lambda|\bar{s} \gamma^\mu b|\Lambda_b\rangle&= \bar
  u_{\Lambda}\big[f_1^V(q^2)\gamma^\mu-f_2^V(q^2)i\sigma^{\mu\nu}\frac{q_\nu}{M_{\Lambda_b}}+f_3^V(q^2)\frac{q^\mu}{M_{\Lambda_b}}\big]
u_{\Lambda_b},\\
 \langle \Lambda|\bar{s} \gamma^\mu\gamma_5 b|\Lambda_b\rangle&= \overline
  u_{\Lambda}\big[f_1^A(q^2)\gamma^\mu-f_2^A(q^2)i\sigma^{\mu\nu}\frac{q_\nu}{M_{\Lambda_b}}+f_3^A(q^2)\frac{q^\mu}{M_{\Lambda_b}}\big]
\gamma_5 u_{\Lambda_b},\\
\langle \Lambda | \bar{s} i\sigma^{\mu q} b | \Lambda_b \rangle &= \bar{u}_\Lambda \left[  \frac{f_1^{TV}(q^2)}{M_{\Lambda_b}} \left(\gamma^\mu q^2 - q^\mu \slashed{q} \right) - f_2^{TV}(q^2) i\sigma^{\mu q}  \right] u_{\Lambda_b}, \\
\langle \Lambda | \bar{s} i\sigma^{\mu q}\gamma_5 b | \Lambda_b \rangle &=\bar{u}_\Lambda \left[  \frac{f_1^{TA}(q^2)}{M_{\Lambda_b}} \left(\gamma^\mu q^2 - q^\mu \slashed{q} \right) - f_2^{TA}(q^2) i\sigma^{\mu q}  \right]\gamma_5 u_{\Lambda_b},        
	    \end{aligned}
	\end{equation}
 where 
 $u_{\Lambda_{(b)}}$ and 
 $M_{\Lambda_{(b)}}$ are the Dirac spinor and mass of $\Lambda _{(b)}$. In turn, we find  that 
	\begin{eqnarray}\label{helicity} 
     H^{Vm}_{\frac{1}{2},\,0} &=&   \sqrt{\frac{Q_-}{q^2}}\left[M_+ F^{Vm}_1(q^2) +\frac{q^2}{M_{\Lambda_b}}
        F^{Vm}_2(q^2)\right]\,, \\
     H^{Vm}_{\frac{1}{2},\, 1} &=&\sqrt{2Q_-}\left[
    F^{Vm}_1(q^2)+\frac{M_+}{M_{\Lambda_b}}
     F^{Vm}_2(q^2)\right]\,,\\
       H^{Vm}_{\frac{1}{2},\,t}&=&   \sqrt{\frac{Q_+}{q^2}}
        \left[M_-  F^{Vm}_1(q^2) +\frac{q^2}
    {M_{\Lambda_b}}  F^{Vm}_3(q^2)\right]\,,\\
     H^{Am}_{\frac{1}{2},\,0} &=&   \sqrt{\frac{Q_+}{q^2}}\left[M_-  F^{Am}_1(q^2) -\frac{q^2}{M_{\Lambda_b}}
        F^{Am}_2(q^2)\right]\,, \\
     H^{Am}_{\frac{1}{2},\, 1} &=&\sqrt{2Q_+}\left[
    F^{Am}_1(q^2)+\frac{M_-}{M_{\Lambda_b}}
     F^{Am}_2(q^2)\right]\,,\\
       H^{Am}_{\frac{1}{2},\,t}&=&   \sqrt{\frac{Q_-}{q^2}}
        \left[M_+  F^{Am}_1(q^2) -\frac{q^2}
    {M_{\Lambda_b}}  F^{Am}_3(q^2)\right]\,,
	\end{eqnarray}
 
 where  
	$M_\pm = M_{\Lambda_b } \pm 
	M_{\Lambda}$, $Q_\pm = (M_\pm )^2 - q^2 $, 
and
\begin{eqnarray}
    F^{V1}_1(q^2)&=&[ C_9^{eff}+C^{\text{NP}}_9 + (C_L + C_R)]f^{V}_1(q^2) - \frac{2m_b}{M_{\Lambda_b}}C_{7\gamma}^{eff}f^{TV}_1(q^2)\,,\\
    F^{V1}_2(q^2)&=&[ C_9^{eff}+C^{\text{NP}}_9 + (C_L + C_R) )]f^{V}_2(q^2) - \frac{2m_b M_{\Lambda_b}}{q^2}C_{7\gamma}^{eff}f^{TV}_2(q^2)\,,\\
    F^{V1}_3(q^2)&=&[ C_9^{eff}+C^{\text{NP}}_9 + (C_L + C_R)]f^{V}_3(q^2)+\frac{2m_b M_-}{q^2}C_{7\gamma}^{eff}f^{TV}_1(q^2)\,,\\
    F^{A1}_1(q^2)&=&[ C_9^{eff}+C^{\text{NP}}_9 - (C_L + C_R)]f^{A}_1(q^2)+\frac{2m_b}{M_{\Lambda_b}}C_{7\gamma}^{eff}f^{TA}_1(q^2)\,,\\
    F^{A1}_2(q^2)&=&[ C_9^{eff}+C^{\text{NP}}_9 - (C_L + C_R) )]f^{A}_2(q^2)+\frac{2m_b M_{\Lambda_b}}{q^2}C_{7\gamma}^{eff}f^{TA}_2(q^2)\,,\\
    F^{A1}_3(q^2)&=&[ C_9^{eff}+C^{\text{NP}}_9 - (C_L + C_R)]f^{V,A}_3(q^2)+\frac{2m_b M_+}{q^2}C_{7\gamma}^{eff}f^{TA}_1(q^2)\,,\\
    F^{V2}_i(q^2)&=&[C_{10} +C^{\text{NP}}_{10} + (C_R - C_L)]f^{V}_i(q^2)\,, \\
    F^{A2}_i(q^2)&=&[C_{10} +C^{\text{NP}}_{10} - (C_R - C_L)]f^{A}_i(q^2)\,, 
\end{eqnarray}
with $i=(1,2,3).$
Combining the relations
$$H^{m}_{\lambda_\Lambda\lambda_m} = H^{Vm}_{\lambda_\Lambda\lambda_m}-H^{Am}_{\lambda_\Lambda\lambda_m}\,,~~~H^{Vm}_{-\lambda_\Lambda,\,-\lambda_m}= H^{Vm}_{\lambda_\Lambda,\, \lambda_m}\,, ~~~H^{Am}_{-\lambda_\Lambda,\,-\lambda_m}= -H^{Am}_{\lambda_\Lambda,\, \lambda_m},$$ 
the evaluations of $H$ are completed once the form factors are given.

	% And have
%  \begin{equation}
%  \label{numernaive t-odd}
%  \begin{aligned}
%     \Delta_{T_1}^+ +\Delta_{T_1}^- &\propto   2 \IM ( a_{+} ^{1} \overline{b_{+} ^{1}} + a_{+} ^{2} \overline{b_{+} ^{2}} - a_{-} ^{1} \overline{b_{-} ^{1}} - a_{-} ^{2} \overline{b_{-} ^{2}} )\\
%     \Delta_{T_1}^+ - \Delta_{T_1}^- &\propto   2 \IM ( a_{+} ^{1} \overline{b_{+} ^{2}} + a_{+} ^{2} \overline{b_{+} ^{1}} - a_{-} ^{1} \overline{b_{-} ^{2}} - a_{-} ^{2} \overline{b_{-} ^{1}} )\\
%      \Delta_{T_1 ^P}^+ + \Delta_{T_1 ^P}^- &\propto  2 \IM ( a_{+} ^{1} \overline{b_{+} ^{1}} + a_{+} ^{2} \overline{b_{+} ^{2}} + a_{-} ^{1} \overline{b_{-} ^{1}} + a_{-} ^{2} \overline{b_{-} ^{2}} )\\
%      \Delta_{T_1 ^P}^+ - \Delta_{T_1 ^P}^- &\propto  2 \IM ( a_{+} ^{1} \overline{b_{+} ^{2}} + a_{+} ^{2} \overline{b_{+} ^{1}} + a_{-} ^{1} \overline{b_{-} ^{2}} + a_{-} ^{2} \overline{b_{-} ^{1}} )\\
% \end{aligned}
%  \end{equation}

	\begin{figure}[tb] 
 \center{\includegraphics[width=15cm]  {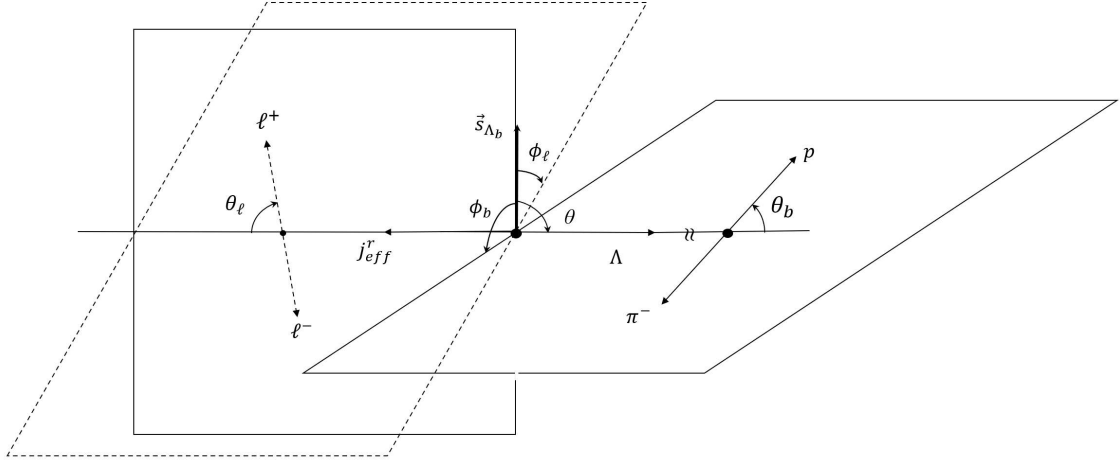}} 
 \caption{\label{plot1} Definitions of the angles }
 \end{figure}

The angular distributions of $\Lambda_b \to \Lambda ( \to p \pi^-) \ell ^+ \ell^-$, related to the kinematic parts, 
are given  by piling $D^J$ to be  
\begin{eqnarray}\label{distru}
&&{\cal D}(q^2,\vec{\Omega})\equiv \frac{\partial^6 \Gamma(\Lambda_b \to \Lambda ( \to p \pi^-) \ell ^+ \ell^-)}{\partial q ^2 \partial \cos \theta \partial \cos \theta_b  \partial \cos \theta_\ell  \partial \phi_b \partial\phi_\ell  } = 
\mathcal{B}(\Lambda \to p \pi^-) \frac{\zeta(q^2)}{32\pi^2} 
\sum_{ \lambda_p \,,\lambda_\pm\,,\lambda_b  }\rho_{\lambda_{\Lambda_b}\lambda_{\Lambda_b }}\left| A_{\lambda_p } \right|^2  \nonumber\\
&&   \left|\sum_m \sum_{\lambda_{m},\lambda_{\Lambda}  } (-1) ^{J_m  }
		H_{\lambda_{\Lambda}\lambda_m }^m
		D^{\frac{1}{2}*}(0,\theta)^{\lambda_b }\, _{\lambda_{\Lambda} - \lambda_m }
		D^{\frac{1}{2}*}(\phi_b,\theta_b)^{\lambda_{\Lambda}}\,_{\lambda_p }
        h^m_{J_m,\lambda_+ \lambda_-}
		D^{J_m *}(\phi_\ell,\theta_\ell )^{\lambda_m }\,_{\lambda_+ -\lambda_-}	 
		\right|^2,\nonumber \\
        &&\zeta(q^2) = \frac{\alpha^2 G_F^2 |V_{ts}^\dagger V_{tb}|^2}{32 \pi^5 } \frac{q^2|\vec{p}_{\Lambda}|}{24M_{\Lambda_b}^2}    \sqrt{1-\delta_\ell},
	\end{eqnarray}
	where $\rho_{\pm,\pm }=(1\pm P_b)/2$,
	$|A_\pm|^2 = (1\pm \alpha) /2$,
	$\lambda_{p} = \pm 1/2$, 
	$|\vec{p}_{\Lambda}|  = \sqrt{Q_+Q_-}/2M_{\Lambda_b}$,  and $J_m=0~(1)$ for $\lambda_m = t~(\pm ,0)$.
The angles are defined in FIG.~\ref{plot1}, where $\theta,\theta_b$ and $\theta_\ell$ are defined in  the CM frames of $\Lambda _{b},\Lambda$ and $\ell^+ \ell^-$, respectively, and $\phi_{b,\ell}$ are the azimuthal angles between the decay planes. 

The physical meaning of Eq.~\eqref{distru} 
is decomposed as follows:
\begin{itemize}
\item The $H^m_{\lambda_\Lambda\lambda_m} D^{\frac{1}{2}*}(0,\theta)^{\lambda_b}\,_{\lambda_\Lambda- \lambda_m}$ is responsible for  $\Lambda_b \to \Lambda j_{eff}^m$, where $H$ and $D$ describe the dynamical and kinematic parts of the amplitudes, respectively. 
\item  The kinematic part of the $\Lambda \to p \pi^-$ is described by  $D^{\frac{1}{2}*}(\phi_b,\theta_b)^{\lambda_{\Lambda}}\,_{\lambda_p }$, while the dynamical part by $|A_{\lambda_p}|$.
\item The terms of $h^m_{J_m,\lambda_+ \lambda_-} $ and $D^{J_m *}(\phi_\ell,\theta_\ell )^{\lambda_m }\,_{\lambda_+ -\lambda_-}	$ describe the dynamical and kinematic parts of  $j^m_{eff} \to \ell^+ \ell^-$, respectively. 
\end{itemize}
The derivation is similar to those of the appendices in Ref.~\cite{TRSEMI}. We cross-check our  results of $\mathcal{D} (\vec{\Omega})$  with Ref.~\cite{Blake:2017angular} and find that  they are matched.
For practical purposes, $\mathcal{D} (\vec{\Omega})$ is expanded as~\cite{LHCb:2018angular} 
 \begin{equation}
 \footnotesize
\begin{aligned}
&{\cal D}(q^2, \vec{\Omega} )  = \frac{3}{32\pi^{2}} \Big(
\left(K_1\sin^2\theta_l+K_2\cos^2\theta_l+K_3\cos\theta_l\right)  +  
 \left(K_4\sin^2\theta_l+K_5\cos^2\theta_l+K_6\cos\theta_l\right)\cos\theta_b +  \\&
 \left(K_7\sin\theta_l\cos\theta_l+K_8\sin\theta_l\right)\sin\theta_b\cos\left(\phi_b+\phi_l\right) +  
\left(K_9\sin\theta_l\cos\theta_l+K_{10}\sin\theta_l\right)\sin\theta_b\sin\left(\phi_b+\phi_l\right) +  \\&
 \left(K_{11}\sin^2\theta_l+K_{12}\cos^2\theta_l+K_{13}\cos\theta_l\right)\cos\theta +  
 \left( K_{14}\sin^2\theta_l+K_{15}\cos^2\theta_l+K_{16}\cos\theta_l\right)\cos\theta_b \cos\theta +  \\&
 \left(K_{17}\sin\theta_l\cos\theta_l+K_{18}\sin\theta_l\right)\sin\theta_b\cos\left(\phi_b+\phi_l\right)\cos\theta  +  
 \left(K_{19}\sin\theta_l\cos\theta_l+K_{20}\sin\theta_l\right)\sin\theta_b\sin\left(\phi_b+\phi_l\right) \cos\theta  \\&
 +\left(K_{21}\cos\theta_l\sin\theta_l+K_{22}\sin\theta_l\right)\sin\phi_l \sin\theta +  
 \left(K_{23}\cos\theta_l\sin\theta_l+K_{24}\sin\theta_l\right)\cos\phi_l  \sin\theta +  \\&
\left(K_{25}\cos\theta_l\sin\theta_l+K_{26}\sin\theta_l\right)\sin\phi_l\cos\theta_b  \sin\theta +  
 \left(K_{27}\cos\theta_l\sin\theta_l+K_{28}\sin\theta_l\right)\cos\phi_l\cos\theta_b  \sin\theta  + \\& 
 \left(K_{29}\cos^2\theta_l+K_{30}\sin^2\theta_l\right)\sin\theta_b\sin\phi_b  \sin\theta +  
 \left(K_{31}\cos^2\theta_l+K_{32}\sin^2\theta_l\right)\sin\theta_b\cos\phi_b  \sin\theta + \\&
 \left(K_{33}\sin^2\theta_l \right) \sin\theta_b\cos\left(2\phi_l+\phi_b\right) \sin\theta  +  
 \left(K_{34}\sin^2\theta_l \right) \sin\theta_b\sin\left(2\phi_l+\phi_b\right)  \sin\theta  \Big)~\,,
\end{aligned}
\label{34terms}
\end{equation} 
where the definitions of $K_i(i=1\sim 34)$ can be found in Appendix~A.
We note that $K_{11\sim 34}$ are proportional to $P_b$, imposing difficulties to extract physical meanings since $P_b$ depends on the productions.
Interestingly, $K_9$ and $K_{10}$ are found to be  
  \begin{equation}
      \begin{aligned}
          K_9 &= \frac{\sqrt{2}\alpha\left(1 - \delta_{\ell}\right)}{4}
         \left(  {\cal T}^R + {\cal T}^L \right)\,,  \\
        K_{10} &= -\frac{\sqrt{2}\alpha\sqrt{1 - \delta_{\ell}}}{4}\left(  {\cal T}^R - {\cal T}^L\right)\,, \\
      \end{aligned}\label{K9K10}
  \end{equation}
  % checked
 which are T-odd  according to Eq.~\eqref{34}. 
We note that  $K_{19,20}$, $K_{21,22}$, $K_{25,26}$, $K_{29,30}$ and $K_{34}$ are also sensitive to the complex phases of NP as they are proportional to the imaginary parts of the  helicity amplitudes.

	\section{Numerical results}
In this work, we estimate the  form factors by the HBM, where the calculation details are given in  Ref.~\cite{centerofmass}. 
The bag parameters adopted in this work are given as 
\begin{equation}
(m_s\,,m_b) = (0.28,4.8)~\text{GeV} \,,~~0.313~\text{GeV}< E_{u,d}<  0.368~\text{GeV}\,, 
\end{equation}
where $R=4.8~\text{GeV}^{-1}$ and $E_q$ are  the bag radius and  quark energy, respectively. 
Recently, $\alpha$ has been updated by BESIII~\cite{bes32018,bes3} with remarkable precision. We take $\alpha =0.732\pm0.014$, $M_{\Lambda_b} = 5.6196$~GeV and the  $\Lambda_b$  lifetime  of $\tau_b=1.471\times 10^{-12}$s from the particle data group~\cite{pdg2022}.
% $\alpha = 0.750$ from bes3, 0.732 from PDG, pdglive
% The asymmetry parameter is taken to be $\alpha =0.732$,  $M_{\Lambda_b}$ and the   $\Lambda_b$  lifetime ($\tau_b$) from Particle Data Group (PDG)~\cite{pdg2022} .
The main uncertainties of the HBM come from $E_q$, affected the form factors largely at  the low $q^2$ region. 
	
  \begin{table}[b]
	\caption{${\cal B}_\ell$ in units of $10^{-6}$} \label{table_compare}
	   \resizebox{\textwidth}{!}{
		\begin{tabular}{lccccccccc}
		\hline
			& HBM & CQM &LCSR&LCSR & BSE&CQM&LCSR&RQM&Data\\	
						& & \cite{gutsche} &\cite{aliev}& \cite{ymwang}& \cite{liu2019}&\cite{mott2015}&\cite{gan2012}&\cite{faustov}&\cite{pdg2022}\\	
			\hline
${\cal B}_e$&0.91(25) &1.0&4.6(1.6)&&$0.660\sim1.208$&&$2.03(^{26}_{9})$&1.07&\multirow{3}{*}{1.08(28)}\\
${\cal B}_\mu $&0.79(18)&1.0&4.0(1.2)&$6.1(^{5.8}_{1.7})$&$0.812\sim1.445$&0.70&&1.05&\\
${\cal B}_\tau $&0.21(2) &0.2&0.8(3)&$2.1(^{2.3}_{0.6})$&$0.252\sim0.392$&0.22&&0.26&\\
			\hline
		\end{tabular} }
\end{table}
 
The total branching fractions are obtained by integrating $\Vec{\Omega}$ and $q^2$ in  Eq.~\eqref{distru}, given as 
\begin{equation}\label{br}
{\cal B}_\ell = {\cal B}(\Lambda_b \to \Lambda \ell^+\ell^-) = \tau_b \int^{M_-^2} _ {4 m_\ell^2}  \zeta ( K_1 + 2 K_2 ) dq^2\,.
\end{equation}
The computed values  and the ones in the literature within the SM are listed in  Table~\ref{table_compare}. In the literature,  Refs.~\cite{gutsche,mott2015} consider the covariant quark model (CQM), Refs.~\cite{aliev,ymwang,gan2012}  light-cone QCD sum rules (LCSR), Ref.~\cite{faustov}  relativistic quark model (RQM),  and Ref.~\cite{liu2019}  Bethe-Salpeter equation (BSE).  We see that 
our results of ${\cal B}_\ell$ are consistent with those of the CQM, RQM and current experimental data but systematically smaller than LCSR. Notably, we find that ${\cal B}_e>{\cal B}_\mu$, which are consistent with Refs.~\cite{faustov} and \cite{aliev}. 
Explicitly, we obtain ${\cal B}_e/{\cal B}_\mu = 1.15$ with little uncertainties due to the correlations.  
The future experiments on ${\cal B}_e/{\cal B}_\mu$ may discriminate the approaches.

Some of the angular observables~($K_i$) bear  special names.
In the following, we  concentrate on $\ell = \mu$. 
 The integrated $ K_{i} $ are defined as 
 \begin{equation}
     \label{averK}
     \langle K_i \rangle = \frac{1}{\Gamma_\kappa} \int^{\kappa'}_{\kappa} \zeta K_i  dq^2 \,,~~~
\Gamma_\kappa = 
\int^{\kappa'} _ {\kappa}  \zeta ( K_1 + 2 K_2 ) dq^2\,. 
 \end{equation}
The  integrated
hadron (lepton)
forward-backward asymmetry of $A_{FB}^h~(A_{FB}^\ell)$ is related to $ \langle K_i \rangle$ through
\begin{eqnarray}\label{hl}
A_{FB}^h = \langle K_4 \rangle + \frac{1}{2} \langle K_5 \rangle \,,~~~A_{FB}^\ell  = \frac{3}{2} \langle  K_3 \rangle \,,
\end{eqnarray}
% \begin{eqnarray}\label{hl}
% A_{FB}^h = \frac{1}{\Gamma_\kappa} \int^{\kappa'}_{\kappa} \zeta \left( K_4 + \frac{1}{2} K_5
% \right)  dq^2 \,,~~~A_{FB}^\ell  = \frac{3}{2\Gamma_\kappa} \int^{\kappa'}_{\kappa} \zeta  K_3  dq^2 \,,
% \end{eqnarray}
while 
\begin{equation}\label{FL}
A_{FB}^{\ell h}  = \frac{3}{4} \langle K_6 \rangle \,,~~~
F_L  = 2 \langle K_1 \rangle - \langle K_2 \rangle  \,,
\end{equation}
% \begin{equation}\label{FL}
% A_{FB}^{\ell h}  = \frac{3}{4\Gamma_\kappa} \int^{\kappa'}_{\kappa} \zeta K_6  dq^2\,,~~~
% F_L  = \frac{1}{\Gamma_\kappa} \int^{\kappa'}_{\kappa} \zeta (2 K_1 - K_2)   dq^2\,,
% \end{equation}
 are the combined forward-backward asymmetry and  longitudinal polarized fraction, respectively. 
The average decay branching fraction is defined as 
\begin{equation}\label{pBR}
\left \langle \frac{ \partial {\cal B}}{\partial q^2} \right  \rangle \equiv \frac{\tau_b}{\kappa' - \kappa}
\Gamma_\kappa\,.
\end{equation}
Note that the $q^2$ regions of $[\kappa,\kappa'] =[8, 11]$ and $[12.5,15]$
in units of GeV$^2$ are contaminated largely by the charmonium resonances, so are not considered.

Our
 results  within the HBM
 are given in Table~\ref{average BR}, along with the ones from the literature and experimental data~\cite{LHCb:exp,LHCb:2018angular}. 
Our values of $A_{FB}^{h,\ell,h\ell}$ and $F_L$ have little uncertainties as  $K_i$ are correlated in the model calculations. 
In the literature, 
 Ref.~\cite{detmold} employs  the lattice QCD, and  Ref.~\cite{faustov} includes the contributions from the charmonium resonances. 
We see that the angular observables in the literature and this work are basically consistent.  
Our  
results of 
$\langle A^h_{FB}\rangle$ and $\langle A^{\ell h}_{FB}\rangle$ are slightly larger than the others due to the updated  $\alpha$\footnote{They used $\alpha=0.642\pm0.013$~\cite{pdg2016}, in sharp contrast to $\alpha=0.732\pm0.014$ adopted in this work.}. Notably, the experimental values of  $A_{FB}^{\ell h}$ are nearly twice larger than the theoretical predictions.

\begin{table}%[hbt!]
\scriptsize
	\caption{ Decay observables, where
$\langle \partial {\cal B} / \partial q^2\rangle $
and 	$\kappa^{(\prime)}$ are  in units of 
$10^{-7}$~GeV$^{-2}$  and 
GeV$^2$, respectively.} 
	\label{average BR}
 \resizebox{0.7\textwidth}{!}{
\begin{tabular}{ccccccc}
\hline
&$[\kappa,\kappa']$&HBM &RQM \cite{faustov}&lattice \cite{detmold}& LHCb \cite{LHCb:exp, LHCb:2018angular}\\
\hline
\multirow{9}{*}{$\left \langle\frac{\partial{\cal B}}{\partial q^2} \right  \rangle$}&$[0.1,2]$&0.25(11)&0.34&0.25(23)&$0.36(^{14}_{13})$\\
&			$[2,4]$&0.16(7) &0.31&0.18(12)&$0.11(^{12}_9)$\\
&			$[4,6]$&  0.20(8)  &0.40&0.23(11)&$0.02(^9_1)$\\
&			$[6,8]$&0.26(9)&0.57&0.307(94)&$0.25(^{13}_{12})$\\
&			$[11,12.5]$&0.44(11)  &0.65&&0.75(21)\\
&			$[15,16]$&0.61(10) &0.72&0.796(75)&1.12(30)\\
&			$[16,18]$&0.65(8) &0.68&0.827(76)&1.22(29)\\
&			$[1.1,6]$& 0.18(7) &0.34&0.20(12)&$0.09(^6_5)$\\
&			$[15,20]$&0.60(6) &0.61&0.756(70)&$1.20(^{26}_{27})$\\
			\hline
\multirow{5}{*}{$A_{FB}^\ell$}&$[0.1,2]$& 0.076(0)&0.067&0.095(15)&$0.37(^{37}_{48})$\\
&			$[11,12.5]$&$-0.357(6)$&$-0.35$&&$0.01(^{20}_{19})$\\
&			$[15,16]$&$-0.403(8)$&$-0.41$&$-0.374(14)$&$-0.10(^{18}_{16})$\\
&			$[16,18]$& $ -0.396(9)$&$-0.36$&$-0.372(13)$&$-0.07(^{14}_{13})$\\
&			$[18,20]$& $ -0.320(9)$&$-0.32$&$-0.309(15)$&$0.01(^{16}_{15})$\\
&			$[15,20]$& $ -0.369(7)$&$-0.33$&$-0.350(13)$&$-0.39(4)$\\
			\hline
\multirow{5}{*}{$A_{FB}^h$}&			$[0.1,2]$&$-0.294(2)$&$-0.26$&$-0.310(18)$&$-0.12(^{34}_{32})$\\
&			$[11,12.5]$&$-0.408(2)$&$-0.30$&&$-0.50(^{11}_{4})$\\
&			$[15,16]$& $-0.384(4)$&$-0.32$&$-0.3069(83)$&$-0.19(^{14}_{16})$\\
&			$[16,18]$&$-0.358(6)$&$-0.31$&$-0.2891(90)$&$-0.44(^{10}_{6})$\\
&			$[18,20]$& $-0.275(6)$&$-0.25$&$-0.227(10)$&$-0.13(^{10}_{12})$\\
&			$[15,20]$& $ -0.333(4)$&$-0.29$&$-0.2710(92)$&$-0.30(5)$\\
			\hline
\multirow{5}{*}{$A_{FB}^{h\ell}$}&			$[0.1,2]$&$-0.028(0)$&$-0.021$&$-0.0302(51)$&\\
&			$[2,4]$&$-0.001(1)$&0.010&$-0.0169(99)$&\\
&			$[4,6]$& 0.047(2)&0.045&0.021(13)&\\
&			$[6,8]$&0.084(1)&0.072&0.053(13)&\\
&			$[15,20]$& 0.179(1)&0.129&0.1398(43)&0.25(4)\\
			\hline
\multirow{5}{*}{$F_L$}&			$[0.1,2]$&0.541(4)&0.66&0.465(84)&$0.56(^{24}_{56})$\\
&			$[11,12.5]$&0.615(0)&0.51&&$0.40(^{37}_{36})$\\
&			$[15,16]$&0.507(1)&0.41&0.454(20)&$0.49(30)$\\
&			$[16,18]$&0.469(0)&0.38&0.417(15)&$0.68(^{15}_{21})$\\
&			$[18,20]$& 0.416(1)&0.35&0.3706(79)&$0.62(^{24}_{27})$\\
			\hline
	\end{tabular}}
\end{table}

After showing that our results in the HBM are compatible with those in the literature, we are ready to estimate the NP contributions to the T-odd observables. From the global fit in the $B$ meson decays~\cite{SinghChundawat:2022zdf}, the permitted imaginary parts of the NP Wilson coefficients  are found  in TABLE~\ref{scen} with four different scenarios\footnote{See FIG.~1 of  Ref.~\cite{SinghChundawat:2022zdf}. It is clear that the signs of NP Wilson coefficients are barely determined. }.
As an illustration, 
we calculate  $\langle K_j \rangle$ with $ K_j \in  \{ K_9,K_{10},K_{19},K_{30} \} $ and 
$(\kappa, \kappa')=( 15~\text{GeV}^2,20~\text{GeV}^2)$  in different scenarios given in
TABLE~\ref{scen}. We fit $P_b$ from the data of  $K_{1-34}$  and find that $P_b$ is  consistent with zero regardless to the presence of  NP.

\begin{table}[h!]
    \caption{ The  Wilson coefficients and  $\langle K_j \rangle$  in units of $10^{-3}$,  with  four NP scenarios. }\label{scen}
    %  with $\kappa = 15(\kappa'= 20)$, in units of $10^{-3}$ and $\text{GeV}^2$, respectively.
    \begin{tabular}{l|cccc|ccccc}\hline 
    Scenarios & $\IM(C_9^{NP})$ &$\IM(C_{10}^{NP})$&$\IM(C_L)$&$\IM (C_R)$& $K_9 $ & $K_{10}$ &$K_{19}$ &$K_{30}$ &$P_b$\\ \hline
    Scenario \#1 & $\pm0.73$&0&0&0&$0$&$\mp 4$&$0$&$0$&$    -0.022( 72)$\\
    Scenario \#2 & 0&$\pm1.86$&0&0\\
    Scenario \#3 & $\pm 1.66$& $\mp 1.66$&0&0&$0$&$\pm 3$&$0$&$0$&$    -0.021(  65)$\\
    Scenario \#4 & $\pm0.77$ &0&$\mp0.77$&$\mp0.77$&$\mp 1$&$\mp 42$&$\mp 1$&$0$&$    -0.019(   64)$\\
    \hline
    \end{tabular}
\end{table}

In the SM, due to lacking of relative complex phases, $\langle K_j \rangle$ are found to be less than $10^{-4}$. Therefore, they provide excellent opportunities to test the SM. 
Since $K_j$ are  proportional to the imaginary parts of the NP Wilson coefficients, which have not been determined yet, their signs remain unknown. However, nonzero values in the experiments would be a smoking gun of NP.   
  Scenario \#1  affects little to $\langle K_j \rangle$, and the results are not listed. 
In addition,  $\langle K_9 \rangle$ is found to be very small  in all the scenarios, which is consistent with the experimental searches. Remarkably, the experimental result of $\langle K_{10} \rangle$ can be explained by Scenario \#4, which can be provided by the $Z'$ model~\cite{Chao:2021qxq,Li:2021cty,Alok:2022pjb}.
The reason can be  traced back to $C_L$ as it interferes largely with the left-handed particles produced by the  SM. 
On the other hand,  $K_{19}$ and $K_{30}$ are highly suppressed by $P_b$.

\section{conclusions}

We have derived the angular distributions of $\Lambda_b \to \Lambda(\to p \pi^-) \ell^+ \ell^-  $ based on the effective schemes of $\Lambda_b \to \Lambda(\to p \pi^-) j_{eff}^m(\to \ell^+ \ell^- ) $. We have shown that our  results are consistent with those in the literature. By studying the effective two-body decays of $\Lambda_b \to \Lambda j_{eff}^m$, we have  
explored the time-reversal asymmetries by
identifying the T-odd correlations in the form of $(\vec{s}_\Lambda \times \vec{s}_m) \cdot \hat{p}$. 
For the numerical estimations, we have adopted the HBM and found that  $\mathcal{B}_e=0.91(25)\times 10^{-6}$, $\mathcal{B}_\mu =0.79(18)\times 10^{-6}$, and  $\mathcal{B}_\tau=0.21(2)\times 10^{-6}$.  For  $\Lambda_b \to \Lambda(\to p \pi^-) \mu^+ \mu^- $,   $A_{FB}^{\ell}$ and $A_{FB}^{h}$, averaged in $15 \leq q^2 \leq 20 \text{GeV}^2$, have been evaluated as  $-0.369(7)$ and $-0.333(4)$, respectively. 
These results are consistent with those in  the literature and experiments, showing that the HBM is suitable for estimating $\Lambda_b \to \Lambda \ell^+\ell^-$. 
We have demonstrated  that $K_9$ and $K_{10}$ are related to $(\vec{s}_\Lambda \times \vec{s}_m) \cdot \hat{p}_\Lambda$,  in which  $K_{10}$ is sensitive to the complex phases generated by NP. 
We have found that $C_L =-0.77i$  can explain the $K_{10}$ puzzle.  We recommend the experiment to revisit $K_{10}$ for a stringent constraint.

	\begin{acknowledgments}
This work is supported in part by the National Key Research and Development Program of China under Grant No. 2020YFC2201501 and  the National Natural Science Foundation of China (NSFC) under Grant No. 12147103.
	\end{acknowledgments}
	
	\newpage
	
	\appendix

        \section{Angular observables}
% Here we take the abbreviations 
%     $a^{1(2)}_{\pm}=H^{1(2)}_{\pm \frac{1}{2},0}$, $b^{1(2)}_{\pm}=H^{1(2)}_{\mp \frac{1}{2},\mp1}$, and $c^{1(2)}_{\pm}=H^{1(2)}_{\pm \frac{1}{2},t}$. 
All $K_i$ are real, which are given as 
        \begin{equation}
        \begin{aligned}
       &K_1 = \frac{1}{4} \Big(- \delta_{\ell} a^{2}_{+} \overline{a^{2}_{+}} - \delta_{\ell} a^{2}_{-} \overline{a^{2}_{-}} + \frac{\delta_{\ell} b^{1}_{+} \overline{b^{1}_{+}}}{2} - \frac{\delta_{\ell} b^{2}_{+} \overline{b^{2}_{+}}}{2} + \frac{\delta_{\ell} b^{1}_{-} \overline{b^{1}_{-}}}{2} - \frac{\delta_{\ell} b^{2}_{-} \overline{b^{2}_{-}}}{2} + \delta_{\ell} c^{2}_{+} \overline{c^{2}_{+}} \\&+ \delta_{\ell} c^{2}_{-} \overline{c^{2}_{-}} + a^{1}_{+} \overline{a^{1}_{+}} + a^{2}_{+} \overline{a^{2}_{+}} + a^{1}_{-} \overline{a^{1}_{-}} + a^{2}_{-} \overline{a^{2}_{-}} + \frac{b^{1}_{+} \overline{b^{1}_{+}}}{2} + \frac{b^{2}_{+} \overline{b^{2}_{+}}}{2} + \frac{b^{1}_{-} \overline{b^{1}_{-}}}{2} + \frac{b^{2}_{-} \overline{b^{2}_{-}}}{2} \Big),\\
       &K_2 =  \frac{1}{4} \Big(\delta_{\ell} a^{1}_{+} \overline{a^{1}_{+}} + \delta_{\ell} a^{1}_{-} \overline{a^{1}_{-}} - \delta_{\ell} b^{2}_{+} \overline{b^{2}_{+}} - \delta_{\ell} b^{2}_{-} \overline{b^{2}_{-}} + \delta_{\ell} c^{2}_{+} \overline{c^{2}_{+}} + \delta_{\ell} c^{2}_{-} \overline{c^{2}_{-}} + b^{1}_{+} \overline{b^{1}_{+}} \\&+ b^{2}_{+} \overline{b^{2}_{+}} + b^{1}_{-} \overline{b^{1}_{-}} + b^{2}_{-} \overline{b^{2}_{-}} \Big),\\
      &K_3 =  - \frac{K_{16}}{P_b} =\frac{ \sqrt{1 - \delta_{\ell}}}{4} \Big(b^{1}_{+}  \overline{b^{2}_{+}} + b^{2}_{+} \overline{b^{1}_{+}} - b^{1}_{-}  \overline{b^{2}_{-}} - b^{2}_{-} \overline{b^{1}_{-}} \Big)\\
        % K_{16}&= -P_b K_3 \\
        &K_4 = \frac{1}{4}\alpha \Big(- \delta_{\ell} a^{2}_{+} \overline{a^{2}_{+}} + \delta_{\ell} a^{2}_{-} \overline{a^{2}_{-}} - \frac{\delta_{\ell} b^{1}_{+} \overline{b^{1}_{+}}}{2} + \frac{\delta_{\ell} b^{2}_{+} \overline{b^{2}_{+}}}{2} + \frac{\delta_{\ell} b^{1}_{-} \overline{b^{1}_{-}}}{2} - \frac{\delta_{\ell} b^{2}_{-} \overline{b^{2}_{-}}}{2} + \delta_{\ell} c^{2}_{+} \overline{c^{2}_{+}} \\&- \delta_{\ell} c^{2}_{-} \overline{c^{2}_{-}} + a^{1}_{+} \overline{a^{1}_{+}} + a^{2}_{+} \overline{a^{2}_{+}} - a^{1}_{-} \overline{a^{1}_{-}} - a^{2}_{-} \overline{a^{2}_{-}} - \frac{b^{1}_{+} \overline{b^{1}_{+}}}{2} - \frac{b^{2}_{+} \overline{b^{2}_{+}}}{2} + \frac{b^{1}_{-} \overline{b^{1}_{-}}}{2} + \frac{b^{2}_{-} \overline{b^{2}_{-}}}{2} \Big),\\
       & K_5 = \frac{1}{4}\alpha \Big(\delta_{\ell} a^{1}_{+} \overline{a^{1}_{+}} - \delta_{\ell} a^{1}_{-} \overline{a^{1}_{-}} + \delta_{\ell} b^{2}_{+} \overline{b^{2}_{+}} - \delta_{\ell} b^{2}_{-} \overline{b^{2}_{-}} + \delta_{\ell} c^{2}_{+} \overline{c^{2}_{+}} - \delta_{\ell} c^{2}_{-} \overline{c^{2}_{-}} - b^{1}_{+} \overline{b^{1}_{+}} \\& - b^{2}_{+} \overline{b^{2}_{+}} + b^{1}_{-} \overline{b^{1}_{-}} + b^{2}_{-} \overline{b^{2}_{-}}\Big),\\
       & K_6 = -\frac{K_{13}}{P_b} =\frac{\alpha\sqrt{1 - \delta_{\ell}}}{4} \Big(- b^{1}_{+}  \overline{b^{2}_{+}} - b^{2}_{+}  \overline{b^{1}_{+}} - b^{1}_{-} \overline{b^{2}_{-}} - b^{2}_{-} \overline{b^{1}_{-}} \Big), \\
            % K_{13} &= -P_b K_6\\
       & K_7 - i K_9 = \frac{\sqrt{2}\alpha\left(1 - \delta_{\ell}\right)}{4} \Big( a^{1}_{-} \overline{b^{1}_{-}} +  a^{2}_{-}  \overline{b^{2}_{-}} -  b^{1}_{+} \overline{a^{1}_{+}} - b^{2}_{+}  \overline{a^{2}_{+}}\Big),\\ 
        % K_7 &= -\frac{\sqrt{2}\alpha\left(1 - \delta_{\ell}\right)}{4}\RE( a^{1}_{-} \overline{b^{1}_{-}} +  a^{2}_{-}  \overline{b^{2}_{-}} -  b^{1}_{+} \overline{a^{1}_{+}} - b^{2}_{+}  \overline{a^{2}_{+}})\\
      &  K_8 - i K_{10} = - \frac{\sqrt{2}\alpha\sqrt{1 - \delta_{\ell}}}{4} \Big(  a^{1}_{-}  \overline{b^{2}_{-}} +  a^{2}_{-}  \overline{b^{1}_{-}} + b^{1}_{+}  \overline{a^{2}_{+}} +  b^{2}_{+}  \overline{a^{1}_{+}} \Big),\\
        % K_8 &= \frac{\sqrt{2}\alpha\sqrt{1 - \delta_{\ell}}}{4}\RE(  a^{1}_{-}  \overline{b^{2}_{-}} +  a^{2}_{-}  \overline{b^{1}_{-}} + b^{1}_{+}  \overline{a^{2}_{+}} +  b^{2}_{+}  \overline{a^{1}_{+}})\\
        % K_9 &= \frac{\sqrt{2}\alpha\left(1 - \delta_{\ell}\right)}{4}\IM( a^{1}_{-} \overline{b^{1}_{-}} +  a^{2}_{-}  \overline{b^{2}_{-}} -  b^{1}_{+} \overline{a^{1}_{+}} - b^{2}_{+}  \overline{a^{2}_{+}})\\
        % K_{10} &= -\frac{\sqrt{2}\alpha\sqrt{1 - \delta_{\ell}}}{4}\IM(  a^{1}_{-}  \overline{b^{2}_{-}} +  a^{2}_{-}  \overline{b^{1}_{-}} + b^{1}_{+}  \overline{a^{2}_{+}} +  b^{2}_{+}  \overline{a^{1}_{+}})\\
      &   K_{11} = \frac{P_b}{4} \Big(- \delta_{\ell} a^{2}_{+} \overline{a^{2}_{+}} + \delta_{\ell} a^{2}_{-} \overline{a^{2}_{-}} + \frac{\delta_{\ell} b^{1}_{+} \overline{b^{1}_{+}}}{2} - \frac{\delta_{\ell} b^{2}_{+} \overline{b^{2}_{+}}}{2} - \frac{\delta_{\ell} b^{1}_{-} \overline{b^{1}_{-}}}{2} + \frac{\delta_{\ell} b^{2}_{-} \overline{b^{2}_{-}}}{2} + \delta_{\ell} c^{2}_{+} \overline{c^{2}_{+}} \\&- \delta_{\ell} c^{2}_{-} \overline{c^{2}_{-}} + a^{1}_{+} \overline{a^{1}_{+}} + a^{2}_{+} \overline{a^{2}_{+}} - a^{1}_{-} \overline{a^{1}_{-}} - a^{2}_{-} \overline{a^{2}_{-}} + \frac{b^{1}_{+} \overline{b^{1}_{+}}}{2} + \frac{b^{2}_{+} \overline{b^{2}_{+}}}{2} - \frac{b^{1}_{-} \overline{b^{1}_{-}}}{2} - \frac{b^{2}_{-} \overline{b^{2}_{-}}}{2} \Big),\\
      &  K_{12} =\frac{P_b}{4} \Big(\delta_{\ell} a^{1}_{+} \overline{a^{1}_{+}} - \delta_{\ell} a^{1}_{-} \overline{a^{1}_{-}} - \delta_{\ell} b^{2}_{+} \overline{b^{2}_{+}} + \delta_{\ell} b^{2}_{-} \overline{b^{2}_{-}} \\&+ \delta_{\ell} c^{2}_{+} \overline{c^{2}_{+}} - \delta_{\ell} c^{2}_{-} \overline{c^{2}_{-}} + b^{1}_{+} \overline{b^{1}_{+}} + b^{2}_{+} \overline{b^{2}_{+}} - b^{1}_{-} \overline{b^{1}_{-}} - b^{2}_{-} \overline{b^{2}_{-}} \Big),\\
        \end{aligned} \label{K1tok10}
        \end{equation}
        \begin{equation}
        \begin{aligned}
       & K_{14} =  \frac{P_b}{4}\alpha \Big(- \delta_{\ell} a^{2}_{+} \overline{a^{2}_{+}} - \delta_{\ell} a^{2}_{-} \overline{a^{2}_{-}} - \frac{\delta_{\ell} b^{1}_{+} \overline{b^{1}_{+}}}{2} + \frac{\delta_{\ell} b^{2}_{+} \overline{b^{2}_{+}}}{2}- \frac{\delta_{\ell} b^{1}_{-} \overline{b^{1}_{-}}}{2} + \frac{\delta_{\ell} b^{2}_{-} \overline{b^{2}_{-}}}{2}+ \delta_{\ell} c^{2}_{+} \overline{c^{2}_{+}} \\&+ \delta_{\ell} c^{2}_{-} \overline{c^{2}_{-}} + a^{1}_{+} \overline{a^{1}_{+}} + a^{2}_{+} \overline{a^{2}_{+}} + a^{1}_{-} \overline{a^{1}_{-}} + a^{2}_{-} \overline{a^{2}_{-}} - \frac{b^{1}_{+} \overline{b^{1}_{+}}}{2} - \frac{b^{2}_{+} \overline{b^{2}_{+}}}{2} - \frac{b^{1}_{-} \overline{b^{1}_{-}}}{2} - \frac{b^{2}_{-} \overline{b^{2}_{-}}}{2} \Big),\\
     &  K_{15} = \frac{P_b}{4}\alpha \Big(\delta_{\ell} a^{1}_{+} \ \overline{a^{1}_{+}} + \delta_{\ell} a^{1}_{-}  \overline{a^{1}_{-}} + \delta_{\ell}  b^{2}_{+} \overline{b^{2}_{+}} + \delta_{\ell}  b^{2}_{-} \overline{b^{2}_{-}} + \delta_{\ell}  c^{2}_{+} \overline{c^{2}_{+}} \\&+ \delta_{\ell}  c^{2}_{-} \overline{c^{2}_{-}} -  b^{1}_{+} \overline{b^{1}_{+}} -  b^{2}_{+} \overline{b^{2}_{+}} -  b^{1}_{-} \overline{b^{1}_{-}} -  b^{2}_{-} \overline{b^{2}_{-}}\Big),\\
       & K_{17} - i K_{19} = -\frac{\sqrt{2}P_b\alpha\left(1 - \delta_{\ell}\right)}{4} \Big( a^{1}_{-} \overline{b^{1}_{-}} +  a^{2}_{-}  \overline{b^{2}_{-}} +  b^{1}_{+} \overline{a^{1}_{+}} + b^{2}_{+}  \overline{a^{2}_{+}}\Big),\\
       &  K_{18} - i K_{20} = -\frac{\sqrt{2}P_b\alpha\sqrt{1 - \delta_{\ell}}}{4} \Big(-  a^{1}_{-}  \overline{b^{2}_{-}} -  a^{2}_{-}  \overline{b^{1}_{-}} + b^{1}_{+}  \overline{a^{2}_{+}} +  b^{2}_{+}  \overline{a^{1}_{+}}\Big),\\
        &   K_{23} - i K_{21} = \frac{P_b\sqrt{2}(1-\delta_{\ell})}{4} \Big( b^{1}_{+} \overline{a^{1}_{-}} -  a^{1}_{+} \overline{b^{1}_{-}} -  a^{2}_{+}  \overline{b^{2}_{-}}  +  b^{2}_{+} \overline{a^{2}_{-}} \Big),\\
         &    K_{24} - i K_{22} = -\frac{P_b\sqrt{2}\sqrt{(1-\delta_{\ell})}}{4} \Big( a^{1}_{+}  \overline{b^{2}_{-}} +  a^{2}_{+}  \overline{b^{1}_{-}} + b^{1}_{+}\overline{a^{2}_{-}} +  b^{2}_{+}  \overline{a^{1}_{-}}\Big),\\
          &    K_{27} - i K_{25} = -\frac{P_b\alpha\sqrt{2}(1-\delta_{\ell})}{4} \Big( -  a^{1}_{+} \overline{b^{1}_{-}} -  a^{2}_{+}  \overline{b^{2}_{-}} -  b^{1}_{+} \overline{a^{1}_{-}} -  b^{2}_{+}  \overline{a^{2}_{-}}\Big),\\
        &     K_{28} - i K_{26} = -\frac{P_b\alpha\sqrt{2}\sqrt{(1-\delta_{\ell})}}{4} \Big(a^{1}_{+}  \overline{b^{2}_{-}} +  a^{2}_{+}  \overline{b^{1}_{-}} - b^{1}_{+}  \overline{a^{2}_{-}} -  b^{2}_{+}\overline{a^{1}_{-}}\Big),\\
          &    K_{31} - i K_{29} = - \frac{P_b\alpha\delta_{\ell}}{2} \Big(  a^{1}_{-} \overline{a^{1}_{+}} +c^{2}_{-} \overline{c^{2}_{+}}\Big),\\
         &      K_{32} - i K_{30} = -\frac{P_b\alpha}{2} \Big( - a^{1}_{-} \overline{a^{1}_{+}} - a^{2}_{-}  \overline{a^{2}_{+}} \Big) +\delta_{\ell} \Big(a^{2}_{-}  \overline{a^{2}_{+}} -c^{2}_{-} \overline{c^{2}_{+}}\Big),\\
         &      K_{33} - i K_{34} =  \frac{P_b\alpha}{4} b^{1}_{+} \overline{b^{1}_{-}}.\\
        \end{aligned} 
        \end{equation}

\end{document}